\documentclass[aps,pre,onecolumn,notitlepage,superscriptaddress,floatfix,letter]{revtex4-1}
\usepackage{graphicx,color}
\usepackage{amsmath,amssymb,bm}
\usepackage{mathtools}
\usepackage{simplewick}
\usepackage{hyperref}
\usepackage{url}
\usepackage{float}

\begin{document}
\title{Signatures of chaos and thermalization\\ in the dynamics of many-body quantum systems}
\author{E. Jonathan Torres-Herrera}
\affiliation{Instituto de F{\'i}sica, Benem\'erita Universidad Aut\'onoma de Puebla, Apt. Postal J-48, Puebla, Puebla, 72570, Mexico}
\author{Lea F. Santos}
\affiliation{Department of Physics, Yeshiva University, New York, New York 10016, USA}
\begin{abstract}
We extend the results of two of our papers [Phys. Rev. A {\bf 94}, 041603R (2016) and Phys. Rev. B {\bf 97}, 060303R (2018)] that touch upon the intimately connected topics of quantum chaos and thermalization. In the first, we argued that when the initial state of isolated lattice many-body quantum systems is chaotic, the power-law decay of the survival probability is caused by the bounds in the spectrum, and thus anticipates thermalization.  In the current work, we provide stronger numerical support for the onset of these algebraic behaviors. In the second paper, we demonstrated that the correlation hole, which is a direct signature of quantum chaos revealed by the evolution of the survival probability at times beyond the power-law decay, appears also for other observables. In the present work, we investigate the correlation hole in the chaotic regime and in the vicinity of a many-body localized phase for the spin density imbalance, which is an observable studied experimentally.
\end{abstract}

\maketitle

\section{Introduction}
\label{intro}
The description of the dynamics of many-body quantum systems perturbed far from equilibrium involves several obstacles. Analytical studies are mostly restricted to integrable models, while the systems are often nonintegrable. Approximations, such as perturbation theory, can capture the behavior of limited time intervals, but different behaviors emerge at different time scales. Furthermore, the dynamics depends on the model, initial state, and observable investigated, which hinders attempts of generalizations.

A useful approach, advocated in Ref.~\cite{Torres2018}, is to study the evolution under full random matrices (FRM), since they allow for the derivation of analytical expressions for any observable at any time. The results can then be used to guide the analysis of realistic chaotic models. FRM from Gaussian orthogonal ensembles (GOEs) are symmetric matrices filled with real random numbers. They were employed by Wigner in the 50's to describe statistically the spectra of heavy nuclei~\cite{Wigner1951P,Wigner1958}. They are unphysical, because they imply the simultaneous interactions of all particles, but they provide relevant information about the level statistics of real systems~\cite{MehtaBook,Guhr1998,StockmannBook}. We have transferred this approach to the study of dynamics. Combining the fact that the eigenstates of FRM are random vectors and that several studies about the correlations among the eigenvalues already exist, one can study the dynamics analytically. Based on the results obtained with FRM, we can then anticipate and justify the behaviors of realistic many-body quantum systems that show GOE level statistics.

The eigenvalues of FRM are correlated; they are prohibited from crossing. Level repulsion is a main signature of quantum chaos. One-body quantum systems that show level repulsion and have a rigid spectrum are known to be chaotic in the classical limit~\cite{StockmannBook}. This has been shown numerically and through semiclassical treatments. As often done, we extend this notion to many-body quantum systems and refer to them as chaotic if they show level repulsion. We note, however, that the relationship between level statistics and classical chaos for many-body quantum systems has not yet been demonstrated. 

In several of our studies of the nonequilibrium dynamics of many-body quantum systems~\cite{Zangara2013,Torres2014PRA,Torres2014PRE,TorresProceed,TorresKollmar2015,Torres2014NJP,
Torres2014PRAb,Torres2015,Torres2016BJP,Torres2016Entropy,Tavora2016,Tavora2017,Torres2017,
Torres2017PTR,CampoARXIV,Torres2018,TorresProceed2017}, we focused on a simple observable, namely the survival probability, which is the probability of finding the system still in its initial state later in time. In hands of a good understanding of the evolution of this quantity, we can then proceed with the analysis of particular observables~\cite{Torres2014NJP,Torres2018}.

The survival probability for GOE-FRM and, equivalently, for chaotic many-body quantum systems shows the following {\em generic} behaviors~\cite{Torres2018}. At short times, the decay is quadratic and universal. Next, the evolution depends on the shape and bounds of the energy distribution of the initial state~\cite{Torres2014PRA,Torres2014PRE,Torres2014NJP,Torres2014PRAb}. In systems perturbed far from equilibrium, where this energy distribution is broad and dense, the decay can initially be very fast, but it subsequently slows down and becomes power law, due to the presence of the unavoidable bounds in the spectrum, which causes the partial reconstruction of the initial state~\cite{Tavora2016,Tavora2017,Khalfin1958,Ersak1969,Fleming1973,Knight1977,Fonda1978,Sluis1991,Urbanowski2009,Campo2011,Campo2016,MugaBook}. Following the algebraic decay, one finds the correlation hole, which is a dip below the saturation of the dynamics~\cite{Torres2017PTR,CampoARXIV,Torres2018,TorresProceed2017} caused by short- and long-correlations between the eigenvalues of chaotic models~\cite{Leviandier1986,Guhr1990,Wilkie1991,Alhassid1992,Gorin2002}. The saturation to a finite value larger than zero  is the last step of the evolution and happens because the systems investigated are finite~\cite{Zangara2013}. 

One needs highly delocalized initial states, large system sizes, and preferably chaotic models to observe the power-law decay of the survival probability induced by the energy bounds. This is because the dynamics needs to detect the energy bounds before detecting the discreteness of the spectrum. Initial states that are so delocalized that most overlaps with the energy eigenbasis are close to random numbers satisfy this criterion. These are chaotic states and they guarantee the onset of thermalization~\cite{Rigol2012,Torres2013,Borgonovi2016}.

The power-law behavior of the survival probability caused by the bounds of the spectrum is evident in FRM of relatively small dimensions, but for realistic models, our numerical results in~\cite{Tavora2016,Tavora2017} were not as convincing. We remedy this situation here by showing results for large system sizes averaged over delocalized initial states and disorder realizations. We consider a one-dimensional (1D) spin-1/2 model with onsite disorder. Averages over initial states and disorder realizations smoothen the curves and reveal the expected algebraic decay.

In Ref.~\cite{Torres2018}, using GOE-FRM, we obtained an expression for the entire evolution of the survival probability and also for the spin density imbalance, which is an observable measured experimentally~\cite{Schreiber2015,Bordia2017}. The results revealed the presence of the correlation hole for both quantities, indicating that the hole is not exclusive to the survival probability. The correlation hole is an unambiguous signature of quantum  chaos and since chaos guarantees that isolated many-body quantum systems reach thermal equilibrium~\cite{Borgonovi2016,Alessio2016}, the presence of the correlation hole also assures thermalization. In the present work, we consider the disordered 1D spin model mentioned above and analyze the correlation hole for the spin density imbalance. We show how the depth of the hole depends on the strength of the disorder and how it vanishes with the approach to a localized phase.

\section{Disordered spin-1/2 model}
\label{sec:Model}

The Hamiltonian of the isolated disordered 1D spin-1/2 model that we study is
\begin{equation}
H = H_0 +V,
\label{Eq:Htotal}
\end{equation}
where 
\begin{eqnarray}
&& H_0= \sum_{k=1}^L h_k  S_k^z     \hspace{0.4 cm} {\rm and} \hspace{0.4 cm}
V=  J \sum_{k=1}^L  \left( S_k^x S_{k+1}^x + S_k^y S_{k+1}^y + S_k^z S_{k+1}^z\right) .
\label{Eq:HoV}
\end{eqnarray}
The chain has $L$ sites and periodic boundary conditions. $S_k^{x,y,z}$ are the spin operators on site $k$, $J$ is the coupling parameter, which we set to 1, and the Zeeman splittings $h_k$ are random numbers from a uniform distribution with support $[-h,h]$, $h$ being the disorder strength. Hamiltonian $H$ conserves the total spin in the $z$-direction, ${\cal S}^z=\sum_kS_k^z$. We focus on the largest subspace, ${\cal S}^z=0$, which has dimension ${\cal D}=L!/(L/2)!^2$.  We list below important features of the disordered spin model, which will be used in the paper.

(i) It represents a many-body quantum system with two-body interactions only, which implies that the density of states is Gaussian~\cite{Brody1981,Kota2001}\footnote{This shape changes in systems with two-body interactions that
  have less than 4 up-spins~\cite{SchiulazARXIV}. }. This is in contrast with the density of states of FRM, which has a semicircular shape~\cite{MehtaBook,Guhr1998,StockmannBook}. 

(ii) When the disorder strength $h$ is of the order of the coupling parameter, $h\sim1$, the eigenvalues of $H$ show strong level repulsion~\cite{Santos2004,SantosEscobar2004,Dukesz2009}. The distribution $P(s)$ of the spacings $s$ between neighboring levels follows the one for GOE-FRM~\cite{Guhr1998},
%\begin{equation}
$P_{GOE}(s)= \frac{\pi}{2} s\exp \left(- \frac{\pi}{4}s^2 \right)$.
%\label{PsFRM}
%\end{equation}
For the system sizes available to exact diagonalization, the best agreement with $P_{GOE}(s)$ happens for $h\sim0.5$ \cite{Torres2017}. 

(iii) As $h$ increases above the coupling parameter, the system moves away from the chaotic region and approaches a localized phase in space~\cite{SantosEscobar2004,Basko2006,Oganesyan2007,Dukesz2009,Torres2015,Torres2017}, despite the presence of interactions. In this limit, the eigenvalues are no longer correlated and  $P(s)$ no longer agrees with $P_{GOE}(s)$. 

\section{Quench Dynamics and Observables}

We prepare the system in the initial state $|\Psi(0) \rangle \equiv |n_0\rangle$, where $|n_0\rangle$ is one of the eigenstates $|n\rangle$ of the initial Hamiltonian $H_0$ [Eq.~\eqref{Eq:HoV}]. These states are site-basis vectors, where on each site there is either a spin pointing up in the $z$-direction or pointing down. The system is then perturbed far from equilibrium and let to evolve under the new total Hamiltonian $H$ [Eq.~\eqref{Eq:Htotal}]. The evolved state is given by
\[
|\Psi(t) \rangle = e^{-i H t} |\Psi(0) \rangle = \sum_{\alpha} C_{n_0}^{\alpha} e^{-i E_{\alpha} t} |\alpha\rangle,
\]
where  $C^{\alpha}_{n_0}= \langle \alpha |  \Psi(0) \rangle $ and $H| \alpha\rangle = E_{\alpha} |\alpha\rangle$.

The two observables that we study are (i) the survival probability and (ii) the spin density imbalance.

(i) The survival probability corresponds to
\begin{equation}
W_{n_0} (t) =\left| \langle \Psi(0) |  \Psi(t) \rangle \right|^2= \left| \sum_{\alpha}  \left| C_{n_0}^{\alpha} \right|^2 e^{-i E_{\alpha}t} \right|^2 .
\label{Eq:SP}
\end{equation}
For very short times, by expanding the exponential, one sees that 
\begin{equation}
W_{n_0}(t)   \approx  1 - \sigma_{n_0}^2 t^2 ,
\label{quadratic}
\end{equation}
where
\begin{equation}
\sigma_{n_0}^2=\sum_{\alpha} |C^{\alpha}_{n_0}|^2 (E_{\alpha} - E_{n_0})^2
\label{stinitialstate}
\end{equation}
is the energy variance of the initial state and
\begin{equation}
E_{n_0}= \langle \Psi (0)| H |\Psi (0)\rangle = \sum_{\alpha} |C^{\alpha}_{n_0}|^2 E_{\alpha}
\label{energyinitialstate}
\end{equation}
is its energy.
The quadratic decay in Eq.~(\ref{quadratic}) is universal and holds when $t \ll \sigma_{n_0}^{ - 1}$.

Equation~(\ref{Eq:SP}) can also be written as
\begin{eqnarray}
W_{n_0}(t) =  \left| \int_{E_{low}}^{E_{up}} dE\, e^{ - iEt} \rho_{n_0}(E) \right|^2,
\label{saitorho}
\end{eqnarray} 
where $E_{low}$ and $E_{up}$ are the lower and upper bounds of the energy spectrum and 
\begin{equation}
\rho_{n_0}(E)  \equiv \sum_{\alpha}  | C^{\alpha}_{n_0} |^2 \delta (E - E_\alpha )
\end{equation} 
is the local density of states (LDOS), that is the energy distribution weighted by the components $| C^{\alpha}_{n_0} |^2$ of the initial state. Therefore, $W_{n_0} (t) $ is  the absolute square of the Fourier transform of the LDOS and $\sigma_{n_0}$ is the width of the LDOS.

(ii) The spin density imbalance~\cite{Luitz2016,Lee2017}, 
\begin{equation}
I(t) = \frac{4}{L} \sum_{k=1}^L S^z_k(0) \langle \Psi(0)| S^z_k(t) |  \Psi(0) \rangle ,
\label{Eq:Imba} 
\end{equation}
is the sum of the correlation in time of the magnetization in the $z$-direction of each site $k$. It is equivalent to the density imbalance  measured  in experiments with cold atoms~\cite{Schreiber2015,Bordia2017}.

\section{Power-law decay}
%\label{sec:1}

Starting the dynamics with an eigenstate of $H_0$ [Eq.~\eqref{Eq:HoV}] and letting it evolve according to $H$ [Eq.~\eqref{Eq:Htotal}] implies that a very strong perturbation is applied to the system. It is equivalent to suddenly changing the coupling parameter $J$ from 0 to 1. In the limit of very strong perturbation, the LDOS has a shape similar to the density of states~\cite{Torres2014NJP}, which in our case, as mentioned in Sec.~\ref{sec:Model}, is Gaussian. If the energy of the initial state is close to the middle of the spectrum, $E_{n_0} \sim 0$, the Gaussian LDOS is nearly symmetric, while the proximity of $E_{n_0}$ to the borders of the spectrum gives rise to skewed distributions~\cite{Torres2014PRAb}. 

The Fourier transform of a Gaussian LDOS leads to the Gaussian decay of the survival probability, $W_{n_0}(t) = \exp(-\sigma_{n_0}^2 t^2)$, as confirmed by our previous works~\cite{Torres2014PRA,Torres2014PRE,TorresProceed,Torres2014NJP,Torres2014PRAb}. However, due to the energy bounds $E_{low}$ and $E_{up}$, the decay is not entirely Gaussian, but given by~\cite{Tavora2016,Tavora2017}
\begin{align}
W_{n_0} (t) &=  \left| \frac{1}{{\cal N} \sqrt {2\pi \sigma_{n_0}^2} }\int_{E_{low}}^{E_{up}}  dE\, e^{ - iEt} e^{ -( E - E_{n_0})^2/2\sigma_{n_0}^2} \right|^2 
\nonumber \\
&
=  \frac{ e^{ - \sigma_{n_0}^2 t^2 } }{4 {\cal N}^2}     \left| \left[   {\rm erf}  \left( \frac{E_{n_0} - E_{low} + i\sigma_{n_0}^2t}{\sqrt{2 \sigma_{n_0}^2} } \right) 
 -   {\rm erf}   \left( \frac{E_{n_0} - E_{up} + i\sigma_{n_0}^2t}{\sqrt{2 \sigma_{n_0}^2} } \right) \right] \right|^2 ,
 \label{Eq:power}
\end{align}
where 
\begin{equation}
{\cal N} = \frac{1}{2} \left[ {\rm erf} \left( \frac{E_{n_0} - E_{low} }{\sqrt{2\sigma_{n_0}^2} } \right) - {\rm erf} \left( \frac{E_{n_0} - E_{up}}{\sqrt {2\sigma_{n_0}^2} } \right) \right]
\end{equation}
is a normalization constant
and $\text{erf}$ is the error function.  In the limit $t \gg \sigma_{n_0}^{-1}$, after averaging out the oscillations, we get
\begin{equation}
W_{n_0}(t \gg \sigma _{n_0}^{ - 1})  \simeq \frac{1}{2\pi {\cal N}^2 \sigma_{n_0}^2 t^2}
\left[ e^{ - (E_{up} - E_{n_0})^2/\sigma_{n_0}^2}  + e^{ - (E_{low} - E_{n_0})^2/\sigma_{n_0}^2}  \right],  
\label{decayline}
\end{equation}
which indicates that a power-law decay $\propto t^{-2}$ should follow the Gaussian behavior. 

In general, the power-law exponent $\gamma$ of $W_{n_0} \propto t^{-\gamma}$ depends on how the LDOS approaches the energy bounds. Asymptotic studies~\cite{Erdelyi1956,Urbanowski2009} have shown that 
\begin{equation}
\gamma = 2 (\xi +1) \hspace{0.3 cm} {\rm for} \hspace{0.3 cm} \rho_{n_0}(E) =(E-E_{low})^{\xi} \eta(E),
\label{Eq:xi}
\end{equation}
where $\lim _{E\rightarrow E_{low}} \eta(E)>0$ (or equivalent for $E_{up}$ if $E_{n_0}$ is closer to the upper bound). In the case of Gaussian or Lorentzian tails, where $\xi \sim 0$, we have that  $\gamma=2$ (see Appendix of~\cite{Tavora2017}).

The results in Eq.~(\ref{Eq:power}) and Eq.~(\ref{Eq:xi}) actually presuppose a continuous spectrum, which is not at all our case. The envelope of our LDOS has a Gaussian shape, but the distribution is obtained from a discrete spectrum. Yet, for enough large ${\cal D}$  and for initial states that are highly delocalized, it takes a long time for the dynamics to resolve the discreteness of the spectrum. Until then, the evolution is equivalent to what we would find in the continuum. 

Highly delocalized initial states appear in the limit of very strong perturbations and for energies $E_{n_0}$ close to the middle of the spectrum. They show up when $H$ (or at least $H_0$) is chaotic. Such chaotic initial states guarantee the onset of thermalization~\cite{Rigol2012,Torres2013,Borgonovi2016}. This led us to claim in Refs.~\cite{Tavora2016,Tavora2017} that in isolated lattice many-body quantum systems with two-body interactions, the emergence of power-law decays given by $W_{n_0}(t)\propto t^{-2}$ indicate that the initial state will eventually reach thermal equilibrium.

\begin{figure}
\resizebox{0.9\columnwidth}{!}{  \includegraphics[height=3cm]{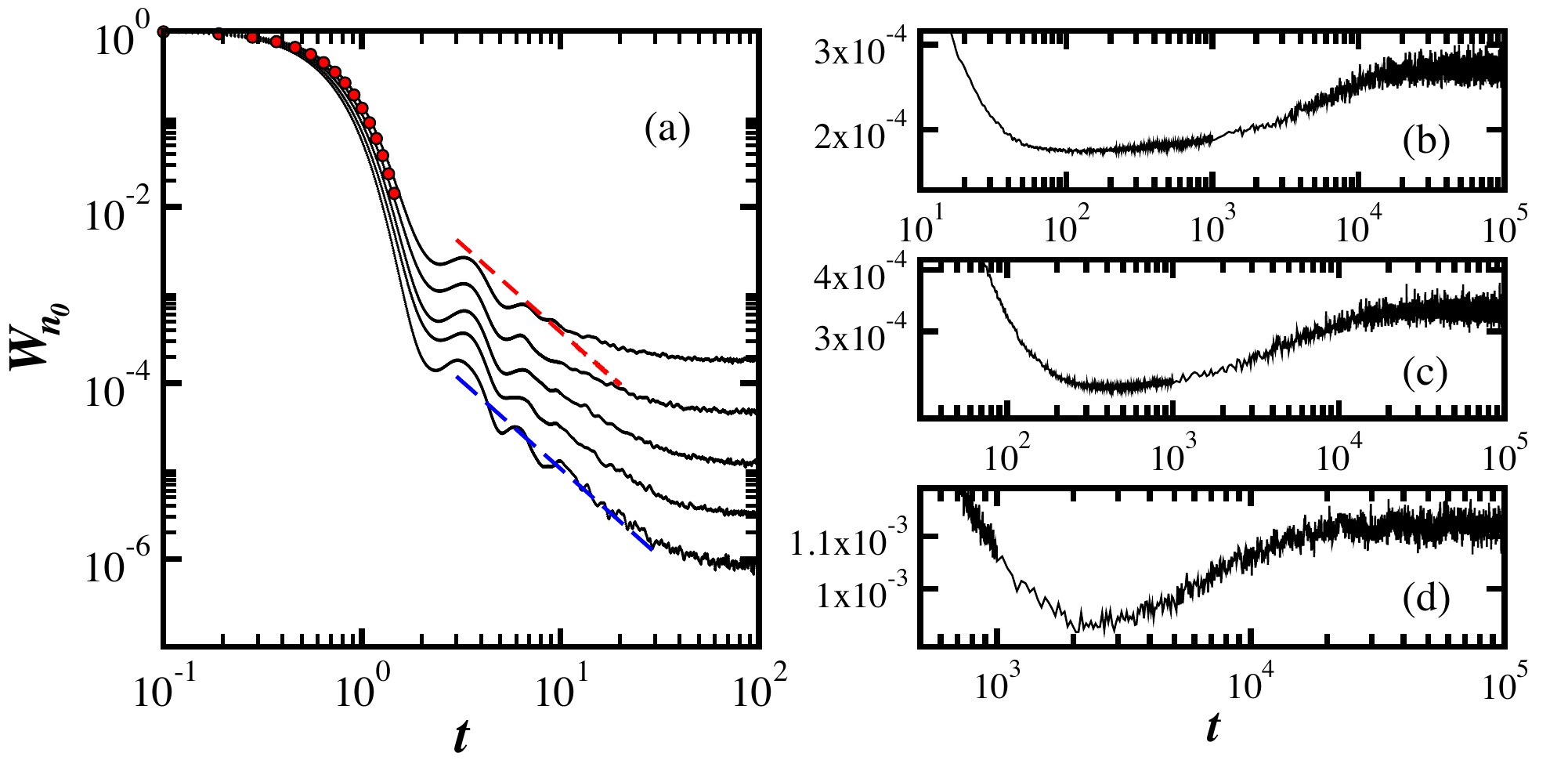} }
\caption{Survival probability for the disordered spin-1/2 model. In (a), the system is in the chaotic regime, $h=0.5$, and the curves from top to bottom are for $L=16, 18, 20, 22, 24$. For $L$ from $16$ to $22$ the average is over $9$ different initial states and over $120$ different disorder realizations. For $L=24$ the average is over $1$ initial state and over $120$ disorder realizations. The top dashed line represents $W_{n_0}(t)\propto t^{-2}$. The bottom dashed line is a fitting for the curve with $L=24$ for $t\in[3,30]$, which gives $W_{n_0}(t)\simeq 0.001 t^{-2}$. The circles are for $W_{n_0}(t) = \exp(-\sigma_{n_0}^2 t^2)$ for $L=16$. On the right panels, we show $W_{n_0}(t)$ for $L=16$ at long times, where the correlation hole emerges. The disorder strengths are $h=0.5$ (b), $h=1$ (c), and $h=1.5$ (c); averages over $1287$ initial states and $77$ disorder realizations. All initial states have $E_{n_0} \sim 0$. }
\label{fig:1}         
\end{figure}

In Fig.~\ref{fig:1} (a), we show $W_{n_0}(t)$ for $H$ in the chaotic limit ($h=0.5$).  System sizes from $L=16$ to $L=24$ are considered. For $L>16$, our computers do not have memory to performing exact diagonalization. To handle such large $L$'s, we use the software package EXPOKIT~\cite{Sidje1998,Expokit}. Since exact diagonalization of $H$ is not available, to guarantee the selection of initial states with  $E_{n_0} \sim 0$ and avoid the borders of the spectrum, we choose initial states with energy $\langle \Psi(0) | H_0 |\Psi(0) \rangle \sim 0$. The curves are smoothened by averages over disorder realizations and different initial states.

In Fig.~\ref{fig:1} (a), the initial decay of $W_{n_0}(t)$ is indeed Gaussian, as anticipated. It is not only the universal quadratic behavior for $t\ll \sigma_{n_0}^{-1}$ presented in Eq.~(\ref{quadratic}), but a true Gaussian decay that holds for times up to $t\sim \sigma_{n_0}^{-1}$. 

The algebraic decay of $W_{n_0}(t)$ dominates after a time $t_P$ obtained from the crossing point between the fast Gaussian decay and the slower power-law behavior~\cite{Tavora2017},
$e^{-\sigma_{n_0}^2 t_P^2} \sim W_{n_0}(t_P \gg \sigma_0^{-1})$.
As seen in  Fig.~\ref{fig:1} (a), after the Gaussian behavior, $W_{n_0}(t)$ decays as $t^{-2}$. Notice that the interval of the power-law decay holds for longer as $L$ increases. This picture is analogous to what one finds for GOE-FRM and implies that equilibration should take longer to happen as the system size increases.

To compare the results above with those for the GOE, we assume that $H_0$ is the diagonal of the FRM and $H$ is the total random matrix. In this case the shape of the envelope of the LDOS is semicircle~\cite{Torres2014PRA,TorresARXIV}, such as the density of states of FRM. The Fourier transform of the semicircle leads to $W_{n_0}(t) = {\cal{J}}_1^2(2\sigma_{n_0}t)/\sigma_{n_0}^2 t^2$, where ${\cal J}_1$ is the Bessel function of the first kind. The Bessel function gives rise to oscillations that decay as $t^{-3}$ (see e.g. Fig.~1 (a) of Ref.~\cite{Torres2018}). The power-law exponent $\gamma=3$ can be justified with Eq.~(\ref{Eq:xi}), using the fact that $\rho_{n_0}(E)$ for FRM approaches the energy bounds as $E^{1/2}$, that is $\xi=1/2$.  Similarly to the disordered spin model, the time interval of the power-law decay of the oscillations under FRM increases with ${\cal D}$, while for very small dimensions, the power-law decay is not observed (see e.g. Fig.~1 (b) of Ref.~\cite{CampoARXIV}). The results for FRM serve as a reference for the analysis of the dynamics of the realistic chaotic spin model. Even if the numerical results for real systems are limited by the accessible system sizes, the claims of the $t^{-2}$ behavior are supported by the theory and by the analogous results obtained with FRM.

\section{Correlation hole for the survival probability}
Realistic systems are finite, so $W_{n_0}(t)$ does not decay to zero. In the absence of too many degeneracies, it saturates to  the infinite time average
\begin{equation}
\overline{W}_{n_0} = \sum_{\alpha} \left| C_{n_0}^{\alpha} \right|^4.
\label{eq:sat}
\end{equation}
$\overline{W}_{n_0}$ measures the level of delocalization of the initial state in the energy eigenbasis. In chaotic initial states, $\overline{W}_{n_0} \propto {\cal D}^{-1}$.

Between the power-law decay and the saturation of the dynamics, one more feature emerges in systems with level repulsion. It corresponds to a dip below the saturation value known as correlation hole. This dip appears only in systems with correlated eigenvalues, being nonexistent in integrable models. It takes a long time to show up, since the dynamics needs to resolve the discreteness of the spectrum.

In Ref.~\cite{Torres2018}, we obtained an expression for the entire evolution of the survival probability evolving under GOE-FRM, which includes the correlation hole. Details can be found in \cite{TorresProceed2017} and \cite{Alhassid1992,MehtaBook}. The basic idea is to write the survival probability in terms of an integral as
\begin{equation}
W_{n_0} (t) =  \int  G(E) e^{-i E t} dE + \overline{W}_{n_0} ,
\end{equation}
where
\begin{equation}
G(E)= \sum _{\alpha_1 \neq \alpha_2} |C^{\alpha_1}_{n_0} |^2 | C^{\alpha_2}_{n_0} |^2 \delta( E - E_{\alpha_1} + E_{\alpha_2}  ) 
\end{equation}
and  $ \overline{W}_{n_0} $ is given by Eq.~(\ref{eq:sat}).  The eigenvalues and eigenstates of FRM are statistically independent. In fact, the eigenstates of FRM are random vectors. This is why one can easily get analytical results using FRM.  The spectral autocorrelation function $G(E)$ can then be separated into $ \langle \sum _{\alpha_1 \neq \alpha_2} |C^{\alpha_1}_{n_0} |^2 | C^{\alpha_2}_{n_0} |^2 \rangle_{\text{FRM}} = 1-\langle \overline{W}_{n_0} \rangle_{\rm{FRM}} $ and $\langle \delta( E - E_{\alpha_1} + E_{\alpha_2}  ) \rangle_{\text{FRM}} $, where $\langle . \rangle_{\rm{FRM}} $ represents the average over ensembles of random matrices. 

The term $\langle \delta( E - E_{\alpha_1} + E_{\alpha_2}  ) \rangle_{\text{FRM}} $ splits into two. One is the Fourier transform of the envelope of the density of states, which leads to a term involving a Bessel function, as described above. The other term detects the discreteness of the spectrum and the correlations between the eigenvalues. It leads to the two-level form factor
\begin{equation}
b_2(\overline{t}) = [1-2\overline{t} + \overline{t} \ln(1+2 \overline{t})] \Theta (1- \overline{t}) 
+ \{-1 + \overline{t} \ln [ (2 \overline{t}+1)/(2 \overline{t} -1) ] \} \Theta(\overline{t}-1),
\label{Eq:hole}
\end{equation}
where $\Theta$ is the Heaviside step function.
Thus, the entire expression for the survival probability is~\cite{Torres2018}
\begin{equation}
\langle W_{n_0} (t) \rangle_{\rm{FRM}}  =  \frac{1-\langle \overline{W}_{n_0} \rangle_{\rm{FRM}} }{{\cal D} -1} \left[ 4 {\cal D} \frac{{\cal J}_1^2 ({\cal E} t)}{({\cal E} t)^2} 
-  b_2 \left( \frac{{\cal E} t}{4 {\cal D} } \right) \right] + \langle \overline{W}_{n_0} \rangle_{\rm{FRM}} ,
\label{Eq:Wo}
\end{equation}
where  $ \pm {\cal E}$ are the energy bounds of the spectrum. The $b_2$ function dominates the dynamics at long times, during the time interval where the correlation hole appears.  This function is directly related with the level number variance~\cite{MehtaBook,Guhr1998,StockmannBook}, which detects long-range correlations between the eigenvalues and is another useful quantity to diagnose quantum chaos. This explains why the hole emerges only in systems with correlated eigenvalues. 

As shown in~\cite{Torres2018}, the $b_2$ function in Eq.~(\ref{Eq:hole}) describes not only the correlation hole for FRM, but also for the chaotic disordered model with $h=0.5$. At such long times, the dynamics depends only on the level of correlations between the eigenvalues and no longer on details about the model and initial state.

On the right panels of Fig.~\ref{fig:1}, we show the correlation hole for different values of the disorder strength: $h=0.5$ (b), $h=1$ (c), and $h=1.5$ (d). The hole fades away as the disorder increases and the system leaves the chaotic region. The hole also gets postponed to later times. The dependence of the time for the correlation hole on the disorder strength and system size, in connection with the notion of Thouless energy, is the subject of a forthcoming paper.

\section{Correlation hole for the spin density imbalance}

In Ref.~\cite{Torres2018}, we also obtained an expression for the entire evolution of the spin density imbalance evolving under FRM, which made evident that the correlation hole is not exclusive to the survival probability. The basic steps for the derivation go as follows. For a generic observable $O$, its dynamics is given by
\begin{equation}
O(t) =  \int K(E) e^{-iEt} dE + \overline{O} ,
\end{equation}
where
\begin{equation}
K(E) =  \sum _{\alpha_1 \neq \alpha_2} C^{\alpha_1}_{n_0}  C^{\alpha_2}_{n_0} O_{\alpha_1 \alpha_2} \delta(E - E_{\alpha_1} + E_{\alpha_2}) ,
\end{equation}
$O_{\alpha_1 \alpha_2} =\langle \alpha_1 | O | \alpha_2 \rangle$, and 
\begin{equation}
\overline{O} = \sum _{\alpha}  |C^{\alpha}_{n_0}|^2 O_{\alpha \alpha}
\label{eq:Odia}
\end{equation}
is the infinite time average. In the FRM model, since the eigenvalues and eigenstates are statistically independent, $K(E)$ splits into 
\begin{align}
\left\langle  \sum _{\alpha_1 \neq \alpha_2} C^{\alpha_1}_{n_0}  C^{\alpha_2}_{n_0} O_{\alpha_1 \alpha_2} \right\rangle_{\text{FRM}} &= \left\langle  \sum _{\alpha_1 , \alpha_2} C^{\alpha_1}_{n_0}  C^{\alpha_2}_{n_0} O_{\alpha_1 \alpha_2} \right\rangle_{\text{FRM}} - \left\langle  \sum _{\alpha} |C^{\alpha}_{n_0}|^2  O_{\alpha \alpha} \right\rangle_{\text{FRM}} \nonumber \\
&= O(0) - \langle \overline{O} \rangle_{\text{FRM}} \nonumber
\end{align}
 and $\langle  \delta(E - E_{\alpha_1} + E_{\alpha_2}) \rangle_{\text{FRM}}$. According to the derivation of Eq.~(\ref{Eq:Wo}) we find that
\begin{align}
\langle I  (t)\rangle_{\text{FRM}} & = \left[ I(0)- \langle \overline{I} \rangle_{\text{FRM}} \right] \left[ \frac{ \langle W_{n_0} (t) \rangle_{\rm{FRM}}  - \langle \overline{W}_{n_0} \rangle_{\rm{FRM}}}{1-\langle \overline{W}_{n_0} \rangle_{\rm{FRM}}}
\right]+ 
\langle \overline{I} \rangle_{\text{FRM}}  
 \label{Eq:Oo} \\
 &= \frac{I(0)- \langle \overline{I} \rangle_{\text{FRM}}}{{\cal D} - 1} \left[ 4 {\cal D} \cfrac{{\cal J}_1^2 (\varepsilon t)}{(\varepsilon t)^2}
 - b_2 \left( \cfrac{\varepsilon t}{4 {\cal D} } \right)\right]  + \langle \overline{I} \rangle_{\text{FRM}}  .
\end{align}

From Eq.~(\ref{Eq:Oo}), it is clear that the evolution of the spin density imbalance is controlled by $\langle W_{n_0} (t) \rangle_{\rm{FRM}}$. The  $t^{-3}$ decay of the oscillations and the onset of the correlation hole, seen for the survival probability, are observed also for $\langle I  (t)\rangle_{\text{FRM}}$ (see the bottom curve of the Fig.~2 of Ref.~\cite{Torres2018}).

\begin{figure}
\resizebox{0.9\columnwidth}{!}{
  \includegraphics[height=3cm]{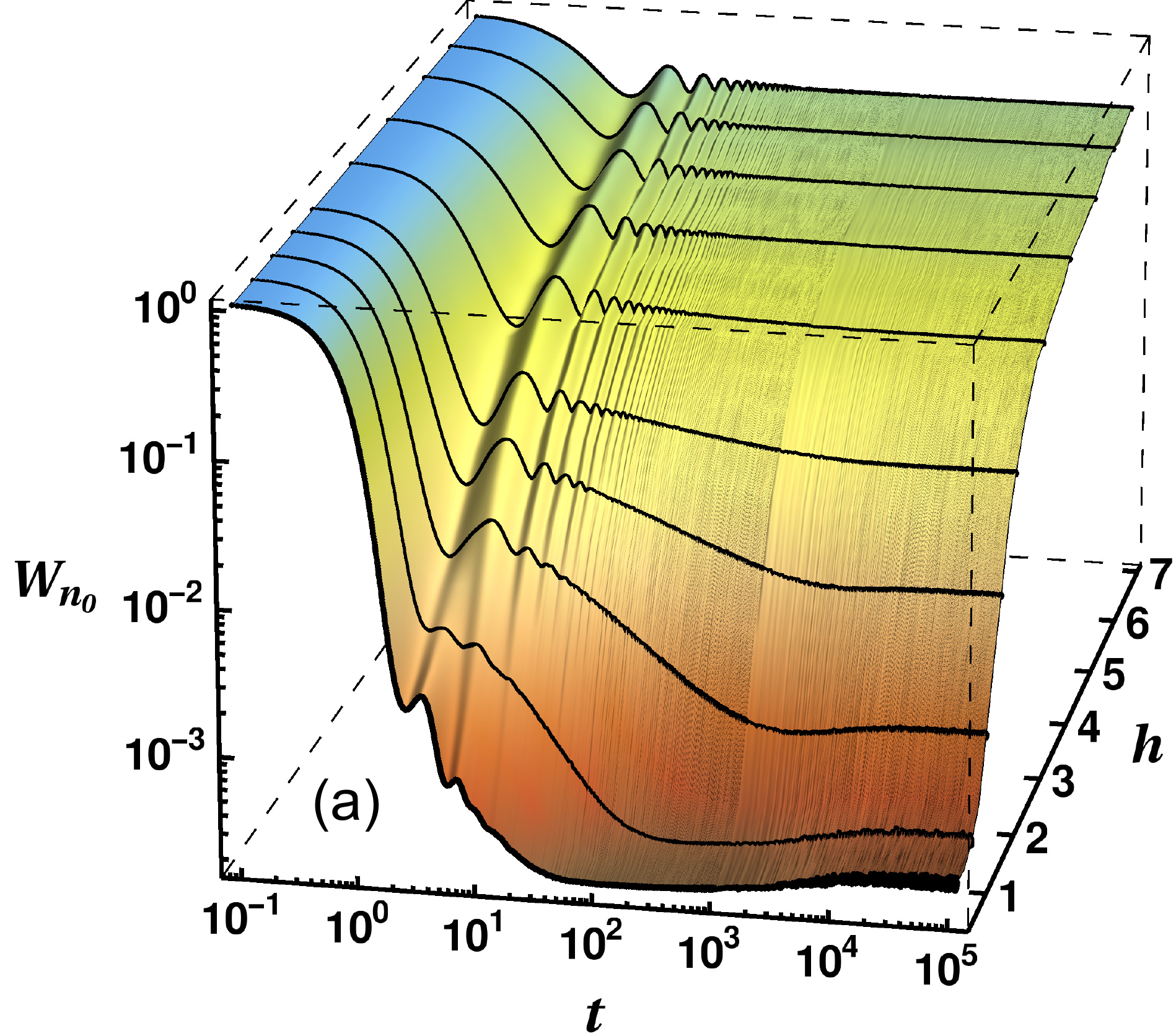} 
\includegraphics[height=3.07cm]{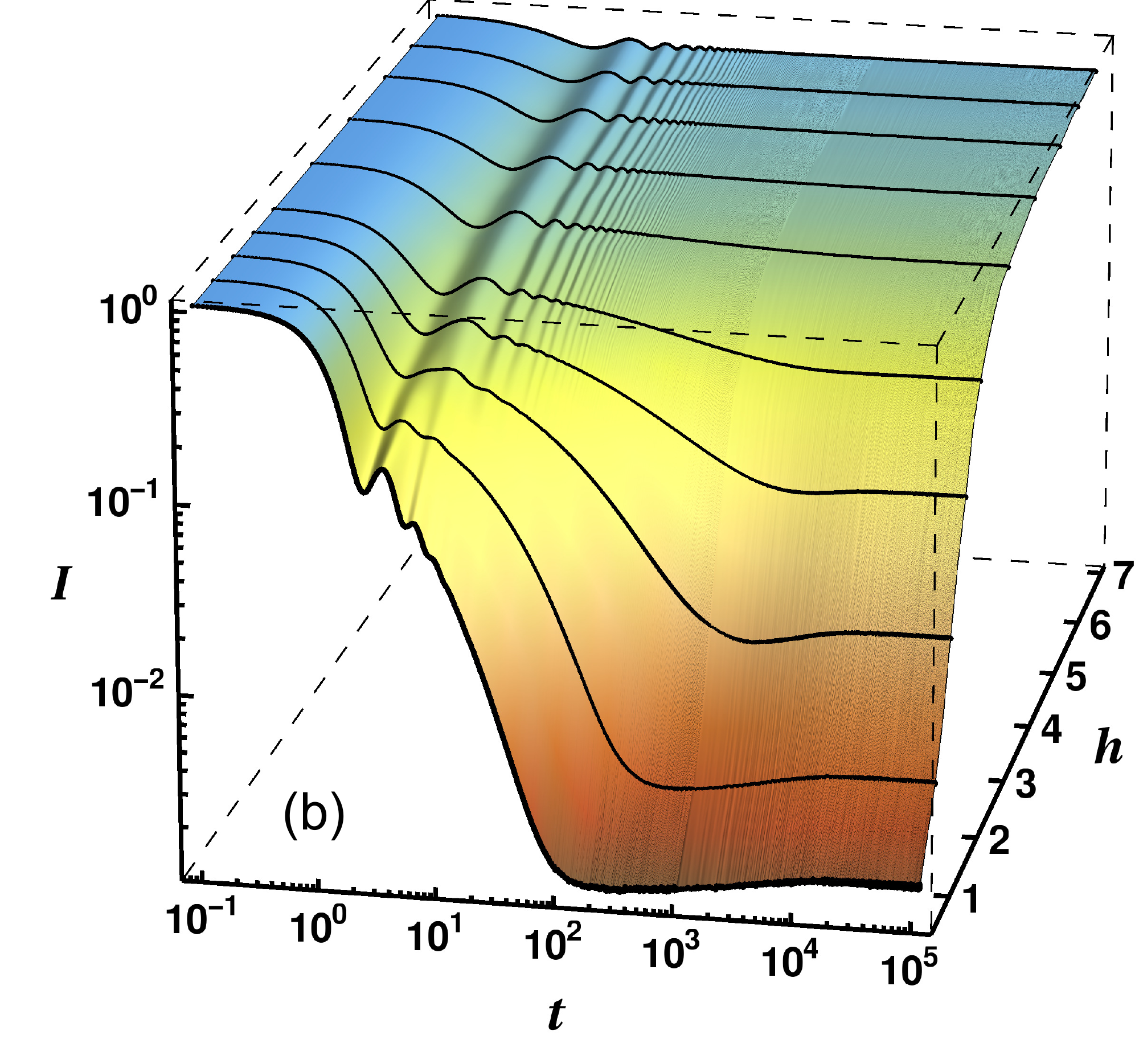} }
\caption{Survival probability and spin density imbalance for the disordered spin-1/2 model for different disorder strengths.  From bottom to top: $h= 0.5, 1.0, 1.5, 2.0, 2.5, 3.5, 4.5, 5.5, 6.25, 7.0$. Average over $1\,287$ initial states with energy close to zero and $77$ disorder realizations.}
\label{fig:2}     
\end{figure}

A similar relationship between $I(t)$ and $W_{n_0}(t)$ is expected also for the disordered spin-1/2 model. As discussed in \cite{Torres2014NJP}, for an observable $O$ that commutes with $H_0$, we have
\begin{equation}
O(t) = O(0) W_{n_0}(t) + \sum_{n\neq n_0} O_{nn}\left| \sum_{\alpha} C_{n}^{\alpha} C_{n_0}^{\alpha} e^{-i E_{\alpha} t} \right|^2 .
\label{Eq:Obs}
\end{equation}
The first term controls the short-time dynamics~\cite{Torres2014NJP}, but at long times, the dynamics depends also on the second term. An expression for $W_{n_0}(t)$ for realistic chaotic systems is available in~\cite{Torres2018}, but further studies are required for obtaining good approximations for the second term. 

The similarity between the behavior of $W_{n_0}(t)$ and $I(t)$ is confirmed in Fig.~\ref{fig:2}. For both quantities the initial decay is Gaussian, the oscillations appear at approximately the same time intervals, the correlation hole for $h=0.5$ is well described~\cite{Torres2018} by the $b_2$ function from Eq.~\eqref{Eq:hole}, and the hole fades away as the disorder strength increases.

\section{Dependence on disorder strength}
In the real system, the results for the power-law decay and the correlation hole for both the survival probability and the spin density imbalance depend on the value of the disorder strength.

\subsection{Power-law decay versus disorder strength}
In  $W_{n_0}(t)$, the source of the power-law decay of the oscillations depends on the disorder strength. When the system is strongly chaotic and the LDOS is ergodically filled, such as when $h\sim0.5$, the power-law decay is caused by the bounds in the spectrum. As the disorder strength increases above $0.5$ and the system approaches spatial localization, the LDOS ceases to be ergodically filled. As shown in Ref.~\cite{Torres2015}, for $h>1$, the power-law decay of $W_{n_0}(t)$ is no longer caused by the energy bounds, but by correlations between the eigenstates, which become fractal. In this scenario, the power-law exponents become smaller than 1. The transition region between the two different power-law behaviors is suggested by the bifurcation seen in Fig.~\ref{fig:2} (a): after the Gaussian decay, the first dip for $h>1$  bifurcates into oscillations for $h<1$. These oscillations must be caused by energy bounds. In contrast, the oscillations seen for $h>1$ must be related with the onset of fractal eigenstates. By comparing Fig.~\ref{fig:2} (a) and Fig.~\ref{fig:2} (b), one sees that the same features  hold also for the spin density imbalance.

\begin{figure}[ht!]
\resizebox{0.9\columnwidth}{!}{  \includegraphics[height=3cm]{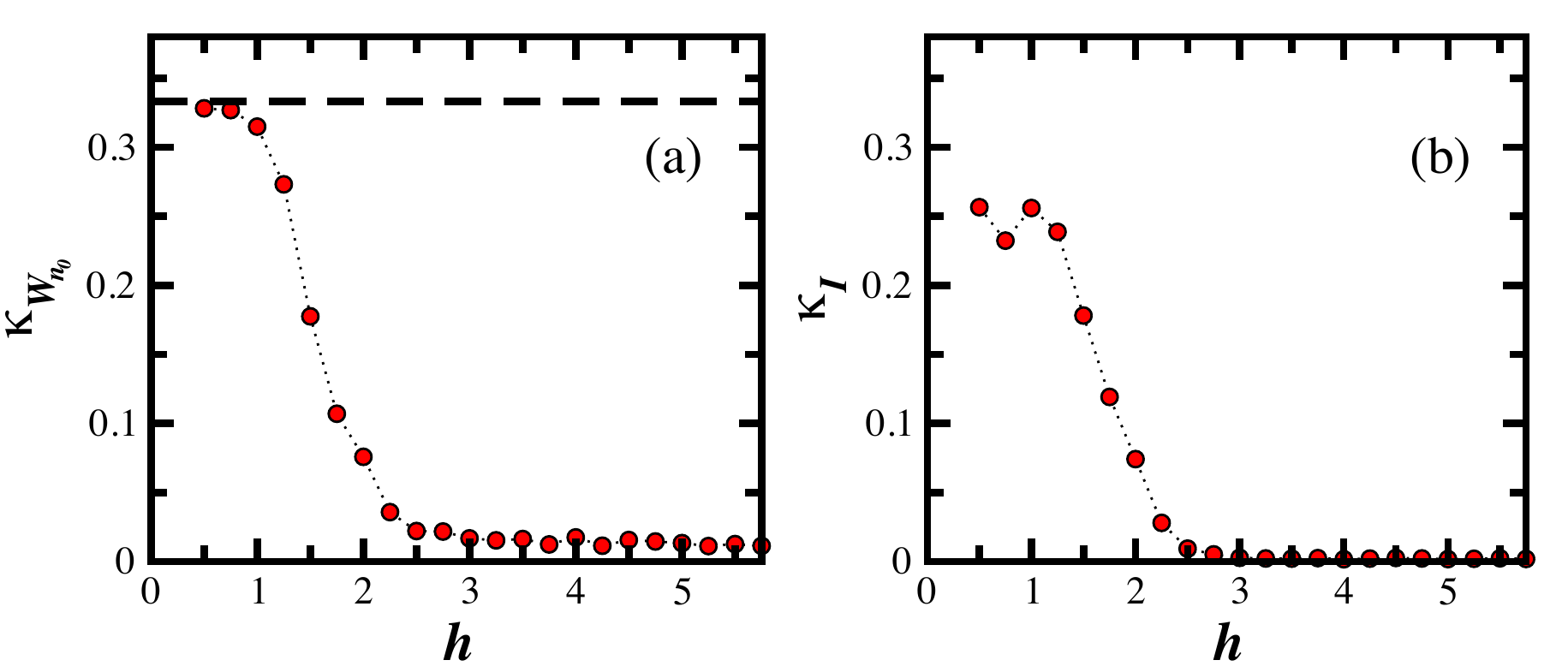} }
\caption{Depth $\kappa$ of the correlation hole for the survival probability (a) and spin density imbalance (b) versus the disorder strength for $h\geq 0.5$. The dashed line indicates the maximum $\left( \kappa_{W_{n_0}}\right)_{\text{FRM}} =1/3$ obtained for the survival probability evolving under FRM. $L=16$, averages over $1\,287$ initial states and $77$ disorder realizations}
\label{fig:3}      
\end{figure}

\subsection{Correlation hole versus disorder strength}
The lowest point of the correlation hole happens at similar times for the survival probability and spin density imbalance. Equivalently to what we did for the survival probability in Ref.~\cite{Torres2017PTR}, we use here the depth of the correlation hole for the density imbalance  to indicate the transition from the chaotic region to the many-body localized phase. We compute
\begin{equation}
\kappa = \frac{ \langle \overline{O} \rangle - \langle O_{min}\rangle}{\langle \overline{O} \rangle  } ,
\end{equation}
where $\langle O_{min}\rangle$ is the minimum value of $\langle O(t) \rangle$. In Fig.~\ref{fig:3} (a), $O = W_{n_0}$ and in Fig.~\ref{fig:3} (b), $O = I$. 

The maximum values of $\kappa_{W_{n_0}}$ and $\kappa_{I}$ happen for FRM. To obtain the infinite time average for GOE-FRM, we use the fact that the coefficients $C_{n_0}^\alpha$ are random numbers from a Gaussian distribution and satisfy the normalization condition $\sum_\alpha|C_{n_0}^\alpha|^2=1$. This implies that
\begin{equation}\label{eq:IPR}
\sum_\alpha |C_{n_0}^\alpha|^4=\frac{3}{{\cal{D}}+2}\quad\Rightarrow\quad\langle \overline{W}_{n_0} \rangle_{\rm{FRM}} =  \frac{3}{{\cal D}+2}
\end{equation}
and
\begin{equation}\label{eq:cross}
\sum_\alpha |C_{n_0}^\alpha|^2 |C_{n}^\alpha|^2=\frac{1}{{\cal{D}}+2}\quad\Rightarrow \quad \langle \overline{I} \rangle_{\rm{FRM}} =  \frac{2}{{\cal D}+2}\,.
\end{equation}
Using Eq.~\eqref{eq:Odia}, the result above is explained as follows:
\begin{equation}
\langle \overline{I} \rangle_{\rm{FRM}} = \sum_{\alpha} |C_{n_0}^\alpha|^2 I_{\alpha \alpha} = \sum_n I_{nn} 
\left( \sum_{\alpha} |C_{n_0}^\alpha|^2 |C_{n}^\alpha|^2 \right) ,
\label{Eq:Inn}
\end{equation}
where
\[
I_{nn} = \frac{4}{L} \sum_k \langle n_0 |S_k^z |n_0 \rangle \langle n |S_k^z |n \rangle .
\]
Among all $I_{nn} $, ${\cal D}/2$ of them equal $+1 $ and ${\cal D}/2$ of them equal $-1$. Thus, they all cancel in the sum in Eq.~(\ref{Eq:Inn}), except for two, because the value of $\sum_{\alpha} |C_{n_0}^\alpha|^2 |C_{n}^\alpha|^2 $ for $n=n_0$ is not the same as for $n\neq n_0$. Since $I_{n_0 n_0} = I(0)=+1$, these two remaining terms give
\begin{equation}
\langle \overline{I} \rangle_{\rm{FRM}} = I(0) \sum_{\alpha} |C_{n_0}^\alpha|^4 - I(0) \sum_{\alpha} |C_{n_0}^\alpha|^2 |C_{n}^\alpha|^2  = \frac{3}{{\cal D}+2} - \frac{1}{{\cal D}+2} = \frac{2}{{\cal D}+2}.
\label{Eq:cancellation}
\end{equation}
In addition to the infinite time averages, we also know that the minimum value of $W_{n_0}(t)$ is $2/{\cal D}$ \cite{Alhassid1992} and for the density imbalance, we found numerically that  $\langle I_{min}\rangle_{\rm{FRM}}=1/{\cal D}$. As a result
\begin{equation}
\left( \kappa_{W_{n_0}} \right)_{\rm{FRM}} =\frac{1}{3} \hspace{0.3 cm} \rm{and} 
\hspace{0.3 cm} \left(\kappa_{I} \right)_{\rm{FRM}} =\frac{1}{2}.
\end{equation}
The equation above implies that for FRM the correlation hole is deeper for the density imbalance than for the survival probability. However, this is not what we observe for real systems [cf. Fig.~\ref{fig:3} (a) and Fig.~\ref{fig:3} (b)]. This suggests that for real systems, the role of the second term in Eq.~(\ref{Eq:Obs}) must be strong and cancellations as those in the derivation of Eq.~(\ref{Eq:cancellation}) must be less likely.

It is clear from Fig.~\ref{fig:3} that the correlation hole detects the transition from chaos to spatial localization. The advantage of having this result also for the density imbalance is that this observable can be studied experimentally.

\section{Conclusions}

Using large system sizes and averages over initial states and disorder realizations, we confirmed the power-law decay $\propto t^{-2}$ for the survival probability of isolated lattice many-body quantum systems that have two-body interactions and are perturbed far from equilibrium. This decay is caused by the presence of bounds in the spectrum. It appears for highly delocalized (chaotic) initial states, which allow for the dynamics to detect the energy bounds before noticing the discreteness of the spectrum.

After the power-law decay, the survival probability of chaotic systems develops a hole before saturating. This correlation hole emerges also in observables, such as the spin density imbalance. We showed how the depth of the correlation hole for this experimental observable fades away as the investigated system leaves the chaotic regime and approaches a localized phase.

As evident from our results, the nonequilibrium dynamics of finite many-body quantum systems shows different behaviors and different time scales. Equilibration of chaotic systems happens only after the power-law decay and the correlation hole.

This paper also mentions some open questions that are part of our current studies. One is the derivation of expressions to describe the evolution of experimental observables under chaotic realistic Hamiltonians. The other is the link between the correlation hole and notions of Thouless energy in disordered systems with interactions.

\begin{acknowledgments}
E.J.T.-H. acknowledges funding from VIEP-BUAP, Mexico. He is also grateful to LNS-BUAP for allowing use of their supercomputing facility. L.F.S. is supported by the NSF grant No. DMR-1603418.
\end{acknowledgments}

%%%%%%%%%%%%%%%%%%%% REFERENCES %%%%%%%%%%%%%%%%%%%%%
%\bibliographystyle{apsrev}
%\bibliography{biblio2018}

\begin{thebibliography}{57}
\expandafter\ifx\csname natexlab\endcsname\relax\def\natexlab#1{#1}\fi
\expandafter\ifx\csname bibnamefont\endcsname\relax
  \def\bibnamefont#1{#1}\fi
\expandafter\ifx\csname bibfnamefont\endcsname\relax
  \def\bibfnamefont#1{#1}\fi
\expandafter\ifx\csname citenamefont\endcsname\relax
  \def\citenamefont#1{#1}\fi
\expandafter\ifx\csname url\endcsname\relax
  \def\url#1{\texttt{#1}}\fi
\expandafter\ifx\csname urlprefix\endcsname\relax\def\urlprefix{URL }\fi
\providecommand{\bibinfo}[2]{#2}
\providecommand{\eprint}[2][]{\url{#2}}

\bibitem[{\citenamefont{Torres-Herrera
  et~al.}(2018)\citenamefont{Torres-Herrera, Garc\'{\i}a-Garc\'{\i}a, and
  Santos}}]{Torres2018}
\bibinfo{author}{\bibfnamefont{E.~J.} \bibnamefont{Torres-Herrera}},
  \bibinfo{author}{\bibfnamefont{A.~M.} \bibnamefont{Garc\'{\i}a-Garc\'{\i}a}},
  \bibnamefont{and} \bibinfo{author}{\bibfnamefont{L.~F.}
  \bibnamefont{Santos}}, \emph{\bibinfo{title}{Generic dynamical features of
  quenched interacting quantum systems: Survival probability, density
  imbalance, and out-of-time-ordered correlator}}, \bibinfo{journal}{Phys. Rev.
  B} \textbf{\bibinfo{volume}{97}}, \bibinfo{pages}{060303R}
  (\bibinfo{year}{2018}).

\bibitem[{\citenamefont{Wigner}(1951)}]{Wigner1951P}
\bibinfo{author}{\bibfnamefont{E.~P.} \bibnamefont{Wigner}},
  \emph{\bibinfo{title}{On the statistical distribution of the widths and
  spacings of nuclear resonance levels}}, \bibinfo{journal}{Proc. Cambridge
  Phil. Soc.} \textbf{\bibinfo{volume}{47}}, \bibinfo{pages}{790}
  (\bibinfo{year}{1951}).

\bibitem[{\citenamefont{Wigner}(1958)}]{Wigner1958}
\bibinfo{author}{\bibfnamefont{E.~P.} \bibnamefont{Wigner}},
  \emph{\bibinfo{title}{On the distribution of the roots of certain symmetric
  matrices}}, \bibinfo{journal}{Ann. Math.} \textbf{\bibinfo{volume}{67}},
  \bibinfo{pages}{325} (\bibinfo{year}{1958}).

\bibitem[{\citenamefont{Mehta}(1991)}]{MehtaBook}
\bibinfo{author}{\bibfnamefont{M.~L.} \bibnamefont{Mehta}},
  \emph{\bibinfo{title}{Random Matrices}} (\bibinfo{publisher}{Academic Press},
  \bibinfo{address}{Boston}, \bibinfo{year}{1991}).

\bibitem[{\citenamefont{Guhr et~al.}(1998)\citenamefont{Guhr,
  M\"uller-Groeling, and Weidenm\"uller}}]{Guhr1998}
\bibinfo{author}{\bibfnamefont{T.}~\bibnamefont{Guhr}},
  \bibinfo{author}{\bibfnamefont{A.}~\bibnamefont{M\"uller-Groeling}},
  \bibnamefont{and} \bibinfo{author}{\bibfnamefont{H.~A.}
  \bibnamefont{Weidenm\"uller}}, \emph{\bibinfo{title}{Random matrix theories
  in quantum physics: Common concepts}}, \bibinfo{journal}{Phys. Rep.}
  \textbf{\bibinfo{volume}{299}}, \bibinfo{pages}{189} (\bibinfo{year}{1998}).

\bibitem[{\citenamefont{St\"ockmann}(2006)}]{StockmannBook}
\bibinfo{author}{\bibfnamefont{H.-J.} \bibnamefont{St\"ockmann}},
  \emph{\bibinfo{title}{Quantum Chaos: An Introduction}}
  (\bibinfo{publisher}{Cambridge University Press},
  \bibinfo{address}{Cambridge}, \bibinfo{year}{2006}).

\bibitem[{\citenamefont{Zangara et~al.}(2013)\citenamefont{Zangara, Dente,
  Torres-Herrera, Pastawski, Iucci, and Santos}}]{Zangara2013}
\bibinfo{author}{\bibfnamefont{P.~R.} \bibnamefont{Zangara}},
  \bibinfo{author}{\bibfnamefont{A.~D.} \bibnamefont{Dente}},
  \bibinfo{author}{\bibfnamefont{E.~J.} \bibnamefont{Torres-Herrera}},
  \bibinfo{author}{\bibfnamefont{H.~M.} \bibnamefont{Pastawski}},
  \bibinfo{author}{\bibfnamefont{A.}~\bibnamefont{Iucci}}, \bibnamefont{and}
  \bibinfo{author}{\bibfnamefont{L.~F.} \bibnamefont{Santos}},
  \emph{\bibinfo{title}{Time fluctuations in isolated quantum systems of
  interacting particles}}, \bibinfo{journal}{Phys. Rev. E}
  \textbf{\bibinfo{volume}{88}}, \bibinfo{pages}{032913}
  (\bibinfo{year}{2013}).

\bibitem[{\citenamefont{Torres-Herrera and
  Santos}(2014{\natexlab{a}})}]{Torres2014PRA}
\bibinfo{author}{\bibfnamefont{E.~J.} \bibnamefont{Torres-Herrera}}
  \bibnamefont{and} \bibinfo{author}{\bibfnamefont{L.~F.}
  \bibnamefont{Santos}}, \emph{\bibinfo{title}{Quench dynamics of isolated
  many-body quantum systems}}, \bibinfo{journal}{Phys. Rev. A}
  \textbf{\bibinfo{volume}{89}}, \bibinfo{pages}{043620}
  (\bibinfo{year}{2014}{\natexlab{a}}).

\bibitem[{\citenamefont{Torres-Herrera and
  Santos}(2014{\natexlab{b}})}]{Torres2014PRE}
\bibinfo{author}{\bibfnamefont{E.~J.} \bibnamefont{Torres-Herrera}}
  \bibnamefont{and} \bibinfo{author}{\bibfnamefont{L.~F.}
  \bibnamefont{Santos}}, \emph{\bibinfo{title}{Local quenches with global
  effects in interacting quantum systems}}, \bibinfo{journal}{Phys. Rev. E}
  \textbf{\bibinfo{volume}{89}}, \bibinfo{pages}{062110}
  (\bibinfo{year}{2014}{\natexlab{b}}).

\bibitem[{\citenamefont{Torres-Herrera and
  Santos}(2014{\natexlab{c}})}]{TorresProceed}
\bibinfo{author}{\bibfnamefont{E.~J.} \bibnamefont{Torres-Herrera}}
  \bibnamefont{and} \bibinfo{author}{\bibfnamefont{L.~F.}
  \bibnamefont{Santos}}, in \emph{\bibinfo{booktitle}{AIP Proceedings}}, edited
  by \bibinfo{editor}{\bibfnamefont{P.}~\bibnamefont{Danielewicz}}
  \bibnamefont{and}
  \bibinfo{editor}{\bibfnamefont{V.}~\bibnamefont{Zelevinsky}}
  (\bibinfo{publisher}{APS}, \bibinfo{address}{East Lansing, Michigan},
  \bibinfo{year}{2014}{\natexlab{c}}).

\bibitem[{\citenamefont{Torres-Herrera
  et~al.}(2015)\citenamefont{Torres-Herrera, Kollmar, and
  Santos}}]{TorresKollmar2015}
\bibinfo{author}{\bibfnamefont{E.~J.} \bibnamefont{Torres-Herrera}},
  \bibinfo{author}{\bibfnamefont{D.}~\bibnamefont{Kollmar}}, \bibnamefont{and}
  \bibinfo{author}{\bibfnamefont{L.~F.} \bibnamefont{Santos}},
  \emph{\bibinfo{title}{Relaxation and thermalization of isolated many-body
  quantum systems}}, \bibinfo{journal}{Phys. Scr. T}
  \textbf{\bibinfo{volume}{165}}, \bibinfo{pages}{014018}
  (\bibinfo{year}{2015}).

\bibitem[{\citenamefont{Torres-Herrera
  et~al.}(2014)\citenamefont{Torres-Herrera, Vyas, and Santos}}]{Torres2014NJP}
\bibinfo{author}{\bibfnamefont{E.~J.} \bibnamefont{Torres-Herrera}},
  \bibinfo{author}{\bibfnamefont{M.}~\bibnamefont{Vyas}}, \bibnamefont{and}
  \bibinfo{author}{\bibfnamefont{L.~F.} \bibnamefont{Santos}},
  \emph{\bibinfo{title}{General features of the relaxation dynamics of
  interacting quantum systems}}, \bibinfo{journal}{New J. Phys.}
  \textbf{\bibinfo{volume}{16}}, \bibinfo{pages}{063010}
  (\bibinfo{year}{2014}).

\bibitem[{\citenamefont{Torres-Herrera and
  Santos}(2014{\natexlab{d}})}]{Torres2014PRAb}
\bibinfo{author}{\bibfnamefont{E.~J.} \bibnamefont{Torres-Herrera}}
  \bibnamefont{and} \bibinfo{author}{\bibfnamefont{L.~F.}
  \bibnamefont{Santos}}, \emph{\bibinfo{title}{Nonexponential fidelity decay in
  isolated interacting quantum systems}}, \bibinfo{journal}{Phys. Rev. A}
  \textbf{\bibinfo{volume}{90}}, \bibinfo{pages}{033623}
  (\bibinfo{year}{2014}{\natexlab{d}}).

\bibitem[{\citenamefont{Torres-Herrera and Santos}(2015)}]{Torres2015}
\bibinfo{author}{\bibfnamefont{E.~J.} \bibnamefont{Torres-Herrera}}
  \bibnamefont{and} \bibinfo{author}{\bibfnamefont{L.~F.}
  \bibnamefont{Santos}}, \emph{\bibinfo{title}{Dynamics at the many-body
  localization transition}}, \bibinfo{journal}{Phys. Rev. B}
  \textbf{\bibinfo{volume}{92}}, \bibinfo{pages}{014208}
  (\bibinfo{year}{2015}).

\bibitem[{\citenamefont{Torres-Herrera
  et~al.}(2016{\natexlab{a}})\citenamefont{Torres-Herrera, T\'avora, and
  Santos}}]{Torres2016BJP}
\bibinfo{author}{\bibfnamefont{E.~J.} \bibnamefont{Torres-Herrera}},
  \bibinfo{author}{\bibfnamefont{M.}~\bibnamefont{T\'avora}}, \bibnamefont{and}
  \bibinfo{author}{\bibfnamefont{L.~F.} \bibnamefont{Santos}},
  \emph{\bibinfo{title}{Survival probability of the n\'eel state in clean and
  disordered systems: an overview}}, \bibinfo{journal}{Braz. J. Phys.}
  \textbf{\bibinfo{volume}{46}}, \bibinfo{pages}{239}
  (\bibinfo{year}{2016}{\natexlab{a}}).

\bibitem[{\citenamefont{Torres-Herrera
  et~al.}(2016{\natexlab{b}})\citenamefont{Torres-Herrera, Karp, T\'avora, and
  Santos}}]{Torres2016Entropy}
\bibinfo{author}{\bibfnamefont{E.~J.} \bibnamefont{Torres-Herrera}},
  \bibinfo{author}{\bibfnamefont{J.}~\bibnamefont{Karp}},
  \bibinfo{author}{\bibfnamefont{M.}~\bibnamefont{T\'avora}}, \bibnamefont{and}
  \bibinfo{author}{\bibfnamefont{L.~F.} \bibnamefont{Santos}},
  \emph{\bibinfo{title}{Realistic many-body quantum systems vs. full random
  matrices: Static and dynamical properties}}, \bibinfo{journal}{Entropy}
  \textbf{\bibinfo{volume}{18}}, \bibinfo{pages}{359}
  (\bibinfo{year}{2016}{\natexlab{b}}).

\bibitem[{\citenamefont{T\'avora et~al.}(2016)\citenamefont{T\'avora,
  Torres-Herrera, and Santos}}]{Tavora2016}
\bibinfo{author}{\bibfnamefont{M.}~\bibnamefont{T\'avora}},
  \bibinfo{author}{\bibfnamefont{E.~J.} \bibnamefont{Torres-Herrera}},
  \bibnamefont{and} \bibinfo{author}{\bibfnamefont{L.~F.}
  \bibnamefont{Santos}}, \emph{\bibinfo{title}{Inevitable power-law behavior of
  isolated many-body quantum systems and how it anticipates thermalization}},
  \bibinfo{journal}{Phys. Rev. A} \textbf{\bibinfo{volume}{94}},
  \bibinfo{pages}{041603R} (\bibinfo{year}{2016}).

\bibitem[{\citenamefont{T\'avora et~al.}(2017)\citenamefont{T\'avora,
  Torres-Herrera, and Santos}}]{Tavora2017}
\bibinfo{author}{\bibfnamefont{M.}~\bibnamefont{T\'avora}},
  \bibinfo{author}{\bibfnamefont{E.~J.} \bibnamefont{Torres-Herrera}},
  \bibnamefont{and} \bibinfo{author}{\bibfnamefont{L.~F.}
  \bibnamefont{Santos}}, \emph{\bibinfo{title}{Power-law decay exponents: A
  dynamical criterion for predicting thermalization}}, \bibinfo{journal}{Phys.
  Rev. A} \textbf{\bibinfo{volume}{95}}, \bibinfo{pages}{013604}
  (\bibinfo{year}{2017}).

\bibitem[{\citenamefont{Torres-Herrera and
  Santos}(2017{\natexlab{a}})}]{Torres2017}
\bibinfo{author}{\bibfnamefont{E.~J.} \bibnamefont{Torres-Herrera}}
  \bibnamefont{and} \bibinfo{author}{\bibfnamefont{L.~F.}
  \bibnamefont{Santos}}, \emph{\bibinfo{title}{Extended nonergodic states in
  disordered many-body quantum systems}}, \bibinfo{journal}{Ann. Phys.
  (Berlin)} \textbf{\bibinfo{volume}{529}}, \bibinfo{pages}{1600284}
  (\bibinfo{year}{2017}{\natexlab{a}}).

\bibitem[{\citenamefont{Torres-Herrera and
  Santos}(2017{\natexlab{b}})}]{Torres2017PTR}
\bibinfo{author}{\bibfnamefont{E.~J.} \bibnamefont{Torres-Herrera}}
  \bibnamefont{and} \bibinfo{author}{\bibfnamefont{L.~F.}
  \bibnamefont{Santos}}, \emph{\bibinfo{title}{Dynamical manifestations of
  quantum chaos: Correlation hole and bulge}}, \bibinfo{journal}{Phil. Trans.
  R. Soc. A} \textbf{\bibinfo{volume}{375}}, \bibinfo{pages}{20160434}
  (\bibinfo{year}{2017}{\natexlab{b}}).

\bibitem[{\citenamefont{del Campo et~al.}()\citenamefont{del Campo,
  Molina-Vilaplana, Santos, and Sonner}}]{CampoARXIV}
\bibinfo{author}{\bibfnamefont{A.}~\bibnamefont{del Campo}},
  \bibinfo{author}{\bibfnamefont{J.}~\bibnamefont{Molina-Vilaplana}},
  \bibinfo{author}{\bibfnamefont{L.~F.} \bibnamefont{Santos}},
  \bibnamefont{and} \bibinfo{author}{\bibfnamefont{J.}~\bibnamefont{Sonner}},
  \emph{\bibinfo{title}{Decay of a thermofield-double state in chaotic quantum
  systems}}, \bibinfo{note}{arXiv:1709.10105}.

\bibitem[{\citenamefont{Santos and Torres-Herrera}(2017)}]{TorresProceed2017}
\bibinfo{author}{\bibfnamefont{L.~F.} \bibnamefont{Santos}} \bibnamefont{and}
  \bibinfo{author}{\bibfnamefont{E.~J.} \bibnamefont{Torres-Herrera}},
  \emph{\bibinfo{title}{Analytical expressions for the evolution of many-body
  quantum systems quenched far from equilibrium}}, \bibinfo{journal}{AIP
  Conference Proceedings} \textbf{\bibinfo{volume}{1912}},
  \bibinfo{pages}{020015} (\bibinfo{year}{2017}).

\bibitem[{\citenamefont{Khalfin}(1958)}]{Khalfin1958}
\bibinfo{author}{\bibfnamefont{L.~A.} \bibnamefont{Khalfin}},
  \emph{\bibinfo{title}{Contribution to the decay theory of a quasi-stationary
  state}}, \bibinfo{journal}{Sov. Phys. JETP} \textbf{\bibinfo{volume}{6}},
  \bibinfo{pages}{1053} (\bibinfo{year}{1958}).

\bibitem[{\citenamefont{Ersak}(1969)}]{Ersak1969}
\bibinfo{author}{\bibfnamefont{L.}~\bibnamefont{Ersak}},
  \emph{\bibinfo{title}{Number of wave functions of an unstable particle}},
  \bibinfo{journal}{Sov. J. Nucl. Phys.} \textbf{\bibinfo{volume}{9}},
  \bibinfo{pages}{263} (\bibinfo{year}{1969}).

\bibitem[{\citenamefont{Fleming}(1973)}]{Fleming1973}
\bibinfo{author}{\bibfnamefont{G.~N.} \bibnamefont{Fleming}},
  \emph{\bibinfo{title}{A unitarity bound on the evolution of nonstationary
  states}}, \bibinfo{journal}{Il Nuovo Cimento} \textbf{\bibinfo{volume}{16}},
  \bibinfo{pages}{232} (\bibinfo{year}{1973}).

\bibitem[{\citenamefont{Knight}(1977)}]{Knight1977}
\bibinfo{author}{\bibfnamefont{P.}~\bibnamefont{Knight}},
  \emph{\bibinfo{title}{Interaction hamiltonians, spectral lineshapes and
  deviations from the exponential decay law at long times}},
  \bibinfo{journal}{Phys. Lett. A} \textbf{\bibinfo{volume}{61}},
  \bibinfo{pages}{25 } (\bibinfo{year}{1977}).

\bibitem[{\citenamefont{Fonda et~al.}(1978)\citenamefont{Fonda, Ghirardi, and
  Rimini}}]{Fonda1978}
\bibinfo{author}{\bibfnamefont{L.}~\bibnamefont{Fonda}},
  \bibinfo{author}{\bibfnamefont{G.~C.} \bibnamefont{Ghirardi}},
  \bibnamefont{and} \bibinfo{author}{\bibfnamefont{A.}~\bibnamefont{Rimini}},
  \emph{\bibinfo{title}{Decay theory of unstable quantum systems}},
  \bibinfo{journal}{Rep. Prog. Phys.,} \textbf{\bibinfo{volume}{41}},
  \bibinfo{pages}{587} (\bibinfo{year}{1978}).

\bibitem[{\citenamefont{Sluis and Gislason}(1991)}]{Sluis1991}
\bibinfo{author}{\bibfnamefont{K.~M.} \bibnamefont{Sluis}} \bibnamefont{and}
  \bibinfo{author}{\bibfnamefont{E.~A.} \bibnamefont{Gislason}},
  \emph{\bibinfo{title}{Decay of a quantum-mechanical state described by a
  truncated lorentzian energy distribution}}, \bibinfo{journal}{Phys. Rev. A}
  \textbf{\bibinfo{volume}{43}}, \bibinfo{pages}{4581} (\bibinfo{year}{1991}).

\bibitem[{\citenamefont{Urbanowski}(2009)}]{Urbanowski2009}
\bibinfo{author}{\bibfnamefont{K.}~\bibnamefont{Urbanowski}},
  \emph{\bibinfo{title}{General properties of the evolution of unstable states
  at long times}}, \bibinfo{journal}{Eur. Phys. J. D}
  \textbf{\bibinfo{volume}{54}}, \bibinfo{pages}{25} (\bibinfo{year}{2009}).

\bibitem[{\citenamefont{del Campo}(2011)}]{Campo2011}
\bibinfo{author}{\bibfnamefont{A.}~\bibnamefont{del Campo}},
  \emph{\bibinfo{title}{Long-time behavior of many-particle quantum decay}},
  \bibinfo{journal}{Phys. Rev. A} \textbf{\bibinfo{volume}{84}},
  \bibinfo{pages}{012113} (\bibinfo{year}{2011}).

\bibitem[{\citenamefont{del Campo}(2016)}]{Campo2016}
\bibinfo{author}{\bibfnamefont{A.}~\bibnamefont{del Campo}},
  \emph{\bibinfo{title}{Exact quantum decay of an interacting many-particle
  system: the calogero-sutherland model}}, \bibinfo{journal}{New J. Phy.}
  \textbf{\bibinfo{volume}{18}}, \bibinfo{pages}{015014}
  (\bibinfo{year}{2016}).

\bibitem[{\citenamefont{Muga et~al.}(2009)\citenamefont{Muga, Ruschhaupt, and
  del Campo}}]{MugaBook}
\bibinfo{author}{\bibfnamefont{J.~G.} \bibnamefont{Muga}},
  \bibinfo{author}{\bibfnamefont{A.}~\bibnamefont{Ruschhaupt}},
  \bibnamefont{and} \bibinfo{author}{\bibfnamefont{A.}~\bibnamefont{del
  Campo}}, \emph{\bibinfo{title}{Time in Quantum Mechanics, vol. 2}}
  (\bibinfo{publisher}{Springer}, \bibinfo{address}{London},
  \bibinfo{year}{2009}).

\bibitem[{\citenamefont{Leviandier et~al.}(1986)\citenamefont{Leviandier,
  Lombardi, Jost, and Pique}}]{Leviandier1986}
\bibinfo{author}{\bibfnamefont{L.}~\bibnamefont{Leviandier}},
  \bibinfo{author}{\bibfnamefont{M.}~\bibnamefont{Lombardi}},
  \bibinfo{author}{\bibfnamefont{R.}~\bibnamefont{Jost}}, \bibnamefont{and}
  \bibinfo{author}{\bibfnamefont{J.~P.} \bibnamefont{Pique}},
  \emph{\bibinfo{title}{Fourier transform: A tool to measure statistical level
  properties in very complex spectra}}, \bibinfo{journal}{Phys. Rev. Lett.}
  \textbf{\bibinfo{volume}{56}}, \bibinfo{pages}{2449} (\bibinfo{year}{1986}).

\bibitem[{\citenamefont{Guhr and Weidenm\"uller}(1990)}]{Guhr1990}
\bibinfo{author}{\bibfnamefont{T.}~\bibnamefont{Guhr}} \bibnamefont{and}
  \bibinfo{author}{\bibfnamefont{H.}~\bibnamefont{Weidenm\"uller}},
  \emph{\bibinfo{title}{Correlations in anticrossing spectra and scattering
  theory. analytical aspects}}, \bibinfo{journal}{Chem. Phys.}
  \textbf{\bibinfo{volume}{146}}, \bibinfo{pages}{21 } (\bibinfo{year}{1990}).

\bibitem[{\citenamefont{Wilkie and Brumer}(1991)}]{Wilkie1991}
\bibinfo{author}{\bibfnamefont{J.}~\bibnamefont{Wilkie}} \bibnamefont{and}
  \bibinfo{author}{\bibfnamefont{P.}~\bibnamefont{Brumer}},
  \emph{\bibinfo{title}{Time-dependent manifestations of quantum chaos}},
  \bibinfo{journal}{Phys. Rev. Lett.} \textbf{\bibinfo{volume}{67}},
  \bibinfo{pages}{1185} (\bibinfo{year}{1991}).

\bibitem[{\citenamefont{Alhassid and Levine}(1992)}]{Alhassid1992}
\bibinfo{author}{\bibfnamefont{Y.}~\bibnamefont{Alhassid}} \bibnamefont{and}
  \bibinfo{author}{\bibfnamefont{R.~D.} \bibnamefont{Levine}},
  \emph{\bibinfo{title}{Spectral autocorrelation function in the statistical
  theory of energy levels}}, \bibinfo{journal}{Phys. Rev. A}
  \textbf{\bibinfo{volume}{46}}, \bibinfo{pages}{4650} (\bibinfo{year}{1992}).

\bibitem[{\citenamefont{Gorin and Seligman}(2002)}]{Gorin2002}
\bibinfo{author}{\bibfnamefont{T.}~\bibnamefont{Gorin}} \bibnamefont{and}
  \bibinfo{author}{\bibfnamefont{T.~H.} \bibnamefont{Seligman}},
  \emph{\bibinfo{title}{Signatures of the correlation hole in total and partial
  cross sections}}, \bibinfo{journal}{Phys. Rev. E}
  \textbf{\bibinfo{volume}{65}}, \bibinfo{pages}{026214}
  (\bibinfo{year}{2002}).

\bibitem[{\citenamefont{Rigol and Srednicki}(2012)}]{Rigol2012}
\bibinfo{author}{\bibfnamefont{M.}~\bibnamefont{Rigol}} \bibnamefont{and}
  \bibinfo{author}{\bibfnamefont{M.}~\bibnamefont{Srednicki}},
  \emph{\bibinfo{title}{Alternatives to eigenstate thermalization}},
  \bibinfo{journal}{Phys. Rev. Lett.} \textbf{\bibinfo{volume}{108}},
  \bibinfo{pages}{110601} (\bibinfo{year}{2012}).

\bibitem[{\citenamefont{Torres-Herrera and Santos}(2013)}]{Torres2013}
\bibinfo{author}{\bibfnamefont{E.~J.} \bibnamefont{Torres-Herrera}}
  \bibnamefont{and} \bibinfo{author}{\bibfnamefont{L.~F.}
  \bibnamefont{Santos}}, \emph{\bibinfo{title}{Effects of the interplay between
  initial state and {H}amiltonian on the thermalization of isolated quantum
  many-body systems}}, \bibinfo{journal}{Phys. Rev. E}
  \textbf{\bibinfo{volume}{88}}, \bibinfo{pages}{042121}
  (\bibinfo{year}{2013}).

\bibitem[{\citenamefont{Borgonovi et~al.}(2016)\citenamefont{Borgonovi,
  Izrailev, Santos, and Zelevinsky}}]{Borgonovi2016}
\bibinfo{author}{\bibfnamefont{F.}~\bibnamefont{Borgonovi}},
  \bibinfo{author}{\bibfnamefont{F.~M.} \bibnamefont{Izrailev}},
  \bibinfo{author}{\bibfnamefont{L.~F.} \bibnamefont{Santos}},
  \bibnamefont{and} \bibinfo{author}{\bibfnamefont{V.~G.}
  \bibnamefont{Zelevinsky}}, \emph{\bibinfo{title}{Quantum chaos and
  thermalization in isolated systems of interacting particles}},
  \bibinfo{journal}{Phys. Rep.} \textbf{\bibinfo{volume}{626}},
  \bibinfo{pages}{1} (\bibinfo{year}{2016}).

\bibitem[{\citenamefont{Schreiber et~al.}(2015)\citenamefont{Schreiber,
  Hodgman, Bordia, L\"uschen, Fischer, Vosk, Altman, Schneider, and
  Bloch}}]{Schreiber2015}
\bibinfo{author}{\bibfnamefont{M.}~\bibnamefont{Schreiber}},
  \bibinfo{author}{\bibfnamefont{S.~S.} \bibnamefont{Hodgman}},
  \bibinfo{author}{\bibfnamefont{P.}~\bibnamefont{Bordia}},
  \bibinfo{author}{\bibfnamefont{H.~P.} \bibnamefont{L\"uschen}},
  \bibinfo{author}{\bibfnamefont{M.~H.} \bibnamefont{Fischer}},
  \bibinfo{author}{\bibfnamefont{R.}~\bibnamefont{Vosk}},
  \bibinfo{author}{\bibfnamefont{E.}~\bibnamefont{Altman}},
  \bibinfo{author}{\bibfnamefont{U.}~\bibnamefont{Schneider}},
  \bibnamefont{and} \bibinfo{author}{\bibfnamefont{I.}~\bibnamefont{Bloch}},
  \emph{\bibinfo{title}{Observation of many-body localization of interacting
  fermions in a quasirandom optical lattice}}, \bibinfo{journal}{Science}
  \textbf{\bibinfo{volume}{349}}, \bibinfo{pages}{842} (\bibinfo{year}{2015}).

\bibitem[{\citenamefont{Bordia et~al.}(2017)\citenamefont{Bordia, L\"uschen,
  Scherg, Gopalakrishnan, Knap, Schneider, and Bloch}}]{Bordia2017}
\bibinfo{author}{\bibfnamefont{P.}~\bibnamefont{Bordia}},
  \bibinfo{author}{\bibfnamefont{H.}~\bibnamefont{L\"uschen}},
  \bibinfo{author}{\bibfnamefont{S.}~\bibnamefont{Scherg}},
  \bibinfo{author}{\bibfnamefont{S.}~\bibnamefont{Gopalakrishnan}},
  \bibinfo{author}{\bibfnamefont{M.}~\bibnamefont{Knap}},
  \bibinfo{author}{\bibfnamefont{U.}~\bibnamefont{Schneider}},
  \bibnamefont{and} \bibinfo{author}{\bibfnamefont{I.}~\bibnamefont{Bloch}},
  \emph{\bibinfo{title}{Probing slow relaxation and many-body localization in
  two-dimensional quasiperiodic systems}}, \bibinfo{journal}{Phys. Rev. X}
  \textbf{\bibinfo{volume}{7}}, \bibinfo{pages}{041047} (\bibinfo{year}{2017}).

\bibitem[{\citenamefont{D'Alessio et~al.}(2016)\citenamefont{D'Alessio, Kafri,
  Polkovnikov, and Rigol}}]{Alessio2016}
\bibinfo{author}{\bibfnamefont{L.}~\bibnamefont{D'Alessio}},
  \bibinfo{author}{\bibfnamefont{Y.}~\bibnamefont{Kafri}},
  \bibinfo{author}{\bibfnamefont{A.}~\bibnamefont{Polkovnikov}},
  \bibnamefont{and} \bibinfo{author}{\bibfnamefont{M.}~\bibnamefont{Rigol}},
  \emph{\bibinfo{title}{From quantum chaos and eigenstate thermalization to
  statistical mechanics and thermodynamics}}, \bibinfo{journal}{Adv. in Phys.}
  \textbf{\bibinfo{volume}{65}}, \bibinfo{pages}{239} (\bibinfo{year}{2016}).

\bibitem[{\citenamefont{Brody et~al.}(1981)\citenamefont{Brody, Flores, French,
  Mello, Pandey, and Wong}}]{Brody1981}
\bibinfo{author}{\bibfnamefont{T.~A.} \bibnamefont{Brody}},
  \bibinfo{author}{\bibfnamefont{J.}~\bibnamefont{Flores}},
  \bibinfo{author}{\bibfnamefont{J.~B.} \bibnamefont{French}},
  \bibinfo{author}{\bibfnamefont{P.~A.} \bibnamefont{Mello}},
  \bibinfo{author}{\bibfnamefont{A.}~\bibnamefont{Pandey}}, \bibnamefont{and}
  \bibinfo{author}{\bibfnamefont{S.~S.~M.} \bibnamefont{Wong}},
  \emph{\bibinfo{title}{Random-matrix physics -- spectrum and strength
  fluctuations}}, \bibinfo{journal}{Rev. Mod. Phys}
  \textbf{\bibinfo{volume}{53}}, \bibinfo{pages}{385} (\bibinfo{year}{1981}).

\bibitem[{\citenamefont{Kota}(2001)}]{Kota2001}
\bibinfo{author}{\bibfnamefont{V.~K.~B.} \bibnamefont{Kota}},
  \emph{\bibinfo{title}{Embedded random matrix ensembles for complexity and
  chaos in finite interacting particle systems}}, \bibinfo{journal}{Phys. Rep.}
  \textbf{\bibinfo{volume}{347}}, \bibinfo{pages}{223} (\bibinfo{year}{2001}).

\bibitem[{\citenamefont{Schiulaz et~al.}()\citenamefont{Schiulaz, T\'avora, and
  Santos}}]{SchiulazARXIV}
\bibinfo{author}{\bibfnamefont{M.}~\bibnamefont{Schiulaz}},
  \bibinfo{author}{\bibfnamefont{M.}~\bibnamefont{T\'avora}}, \bibnamefont{and}
  \bibinfo{author}{\bibfnamefont{L.~F.} \bibnamefont{Santos}},
  \emph{\bibinfo{title}{From few- to many-body quantum systems}},
  \bibinfo{note}{arXiv:1802.08691}.

\bibitem[{\citenamefont{Santos}(2004)}]{Santos2004}
\bibinfo{author}{\bibfnamefont{L.~F.} \bibnamefont{Santos}},
  \emph{\bibinfo{title}{Integrability of a disordered {H}eisenberg spin-1/2
  chain}}, \bibinfo{journal}{J. Phys. A} \textbf{\bibinfo{volume}{37}},
  \bibinfo{pages}{4723} (\bibinfo{year}{2004}).

\bibitem[{\citenamefont{Santos et~al.}(2004)\citenamefont{Santos, Rigolin, and
  Escobar}}]{SantosEscobar2004}
\bibinfo{author}{\bibfnamefont{L.~F.} \bibnamefont{Santos}},
  \bibinfo{author}{\bibfnamefont{G.}~\bibnamefont{Rigolin}}, \bibnamefont{and}
  \bibinfo{author}{\bibfnamefont{C.~O.} \bibnamefont{Escobar}},
  \emph{\bibinfo{title}{Entanglement versus chaos in disordered spin systems}},
  \bibinfo{journal}{Phys. Rev. A} \textbf{\bibinfo{volume}{69}},
  \bibinfo{pages}{042304} (\bibinfo{year}{2004}).

\bibitem[{\citenamefont{Dukesz et~al.}(2009)\citenamefont{Dukesz, Zilbergerts,
  and Santos}}]{Dukesz2009}
\bibinfo{author}{\bibfnamefont{F.}~\bibnamefont{Dukesz}},
  \bibinfo{author}{\bibfnamefont{M.}~\bibnamefont{Zilbergerts}},
  \bibnamefont{and} \bibinfo{author}{\bibfnamefont{L.~F.}
  \bibnamefont{Santos}}, \emph{\bibinfo{title}{Interplay between interaction
  and (un)correlated disorder in one-dimensional many-particle systems:
  delocalization and global entanglement}}, \bibinfo{journal}{New J. Phys.}
  \textbf{\bibinfo{volume}{11}}, \bibinfo{pages}{043026}
  (\bibinfo{year}{2009}).

\bibitem[{\citenamefont{Basko et~al.}(2006)\citenamefont{Basko, Aleiner, and
  Altshuler}}]{Basko2006}
\bibinfo{author}{\bibfnamefont{D.~M.} \bibnamefont{Basko}},
  \bibinfo{author}{\bibfnamefont{I.~L.} \bibnamefont{Aleiner}},
  \bibnamefont{and} \bibinfo{author}{\bibfnamefont{B.~L.}
  \bibnamefont{Altshuler}}, \emph{\bibinfo{title}{Metal-insulator transition in
  a weakly interacting many-electron system with localized single-particle
  states}}, \bibinfo{journal}{Ann. Phys.} \textbf{\bibinfo{volume}{321}},
  \bibinfo{pages}{1126} (\bibinfo{year}{2006}).

\bibitem[{\citenamefont{Oganesyan and Huse}(2007)}]{Oganesyan2007}
\bibinfo{author}{\bibfnamefont{V.}~\bibnamefont{Oganesyan}} \bibnamefont{and}
  \bibinfo{author}{\bibfnamefont{D.~A.} \bibnamefont{Huse}},
  \emph{\bibinfo{title}{Localization of interacting fermions at high
  temperature}}, \bibinfo{journal}{Phys. Rev. B} \textbf{\bibinfo{volume}{75}},
  \bibinfo{pages}{155111} (\bibinfo{year}{2007}).

\bibitem[{\citenamefont{Luitz et~al.}(2016)\citenamefont{Luitz, Laflorencie,
  and Alet}}]{Luitz2016}
\bibinfo{author}{\bibfnamefont{D.~J.} \bibnamefont{Luitz}},
  \bibinfo{author}{\bibfnamefont{N.}~\bibnamefont{Laflorencie}},
  \bibnamefont{and} \bibinfo{author}{\bibfnamefont{F.}~\bibnamefont{Alet}},
  \emph{\bibinfo{title}{Extended slow dynamical regime close to the many-body
  localization transition}}, \bibinfo{journal}{Phys. Rev. B}
  \textbf{\bibinfo{volume}{93}}, \bibinfo{pages}{060201}
  (\bibinfo{year}{2016}).

\bibitem[{\citenamefont{Lee et~al.}(2017)\citenamefont{Lee, Look, Lim, and
  Sheng}}]{Lee2017}
\bibinfo{author}{\bibfnamefont{M.}~\bibnamefont{Lee}},
  \bibinfo{author}{\bibfnamefont{T.~R.} \bibnamefont{Look}},
  \bibinfo{author}{\bibfnamefont{S.~P.} \bibnamefont{Lim}}, \bibnamefont{and}
  \bibinfo{author}{\bibfnamefont{D.~N.} \bibnamefont{Sheng}},
  \emph{\bibinfo{title}{Many-body localization in spin chain systems with
  quasiperiodic fields}}, \bibinfo{journal}{Phys. Rev. B}
  \textbf{\bibinfo{volume}{96}}, \bibinfo{pages}{075146}
  (\bibinfo{year}{2017}).

\bibitem[{\citenamefont{Erd\'elyi}(1956)}]{Erdelyi1956}
\bibinfo{author}{\bibfnamefont{A.}~\bibnamefont{Erd\'elyi}},
  \emph{\bibinfo{title}{Asymptotic expansions of fourier integrals involving
  logarithmic singularities}}, \bibinfo{journal}{J. Soc. Indust. Appr. Math.}
  \textbf{\bibinfo{volume}{4}}, \bibinfo{pages}{38} (\bibinfo{year}{1956}).

\bibitem[{\citenamefont{Sidje}(1998)}]{Sidje1998}
\bibinfo{author}{\bibfnamefont{R.~B.} \bibnamefont{Sidje}},
  \emph{\bibinfo{title}{Expokit: A software package for computing matrix
  exponentials}}, \bibinfo{journal}{ACM Trans. Math. Softw.}
  \textbf{\bibinfo{volume}{24}}, \bibinfo{pages}{130} (\bibinfo{year}{1998}).

\bibitem[{Exp()}]{Expokit}
\emph{\bibinfo{title}{{Expokit}}},
  \bibinfo{howpublished}{\url{http://www.maths.uq.edu.au/expokit/}}.

\bibitem[{\citenamefont{Santos and Torres-Herrera}()}]{TorresARXIV}
\bibinfo{author}{\bibfnamefont{L.~F.} \bibnamefont{Santos}} \bibnamefont{and}
  \bibinfo{author}{\bibfnamefont{E.~J.} \bibnamefont{Torres-Herrera}},
  \emph{\bibinfo{title}{Nonequilibrium many-body quantum dynamics: from full
  random matrices to real systems}}, \bibinfo{note}{arXiv:1803.06012}.

\end{thebibliography}

\end{document}